# ICT in Local Self Governance: A Study of Rural India

Puneet Kumar
Assistant Professor
MITS University
Laxmangarh, Rajasthan

Dharminder Kumar
Professor & Chairman
Department of CSE
GJUST, Hisar, Haryana

Narendra Kumar
Assistant Professor
MITS University
Laxmangarh, Rajasthan

## ABSTRACT
The concept of local self-governance is not new as it has its roots in ancient time even before the era of Mauryan emperors. This paper depicts the journey of local self-governance from antediluvian time to 21st century. Further, in the current scenario Information and Communication Technology (ICT) has emerged as a successful tool for dissemination of various e-governance services and in this regard the Government of India has formulated NeGP with adequate service delivery mechanism. With the inculcation of ICT, various applications were designed by central as well as state governments which lead towards strengthening of PRIs for rural reform. This paper also throws some light on necessity of ICT in self-governance along with some case studies.

## 1. INTRODUCTION
The democratization in Indian government system is stances on the edges of local self-governance. It intends involvement and participation of people at every level. Such an architecture of government system termed as by the people, of the people and for the people and it undergoes devolution of powers, responsibilities and functions thus forming a largest democracy ubiquitously. The rural population represents two third of Indian population [1] therefore their participation and involvement is more necessary in every aspect of governance. India endorses a bottom approach of governance i.e. a village panchayat and ward is treated as a primary unit of governance in context of rural and urban areas respectively [2]. According to 73rd constitutional amendment, three tiers architecture for the rural local self-governance has been formulated comprises of Gram panchayat at village level, Panchayat samiti or Taluk panchayat at block level and Zila parishad or Zila panchayat at district level commonly referred as Panchayati Raj Institutions (PRIs) [3][4]. Since every tier of PRIs encompasses funds, functions and functionaries [5] associated with it and moreover the literacy rate is quite low in comparison with urban areas; therefore they need to be managed keenly. With the advent of Information and Communication Technology (ICT) various e-governance applications are devised for rural development and management of PRIs comprises of various aspects like finances, accounts, procurement, land records, agriculture marketing etc. In this regard various pilot projects have been started across the whole nation to accomplish diversified requirements thereafter the successful e-governance projects are generalized for the whole community. This paper discusses some landmark projects and throws a spotlight on current status of e-governance in rural areas and along with major impediments in their implementation.

## 2. HISTORY AND EVOLUTION OF LOCAL SELF-GOVERNANCE IN INDIA
The administration in the villages by the local people has its origin in the period 236 B.C. – 324 B.C. when size of villages varied from 100 to 500 families and their boundaries were demarcated on the basis of natural resources like rivers, lakes, ponds etc. In that era the village administration staff comprises of Headman (Adhyaksha), accountant (Samkhayaka), other officials (Sthanikas), medical officer (Chikitska) etc. with the objective of serving the whole community. But this village assembly has lost his powers and utility in the regime of Mauryan emperors because of their focus on centralized administration. However for resolution of conflicts and other matter of disputes, the village assembly comprises of elder people of village. These village assemblies or Panchayats were also active in the era of Harshavardhan in 16th century. The system was in existence at the time of Gupt emperors with some changes in jargon. The villages were administered by Grampati assisted by village assemblies. Further, these village assemblies were fully competent in acquisition and sale of land, utilization of common land for revenue generation and other purposes. The era of Mughals was having despotism to great extent and the will of King (Sultan) was the law; as a result local self-governing institutions were became deprived. Their roles were confined towards managing their own local affairs. With the arrival of Muslims after conquering Sindh & Multan in 712 A.D. the situation of village assemblies became pathetic and their independence began to diminish. The Mughal emperors were keenly interested in revenue generation by various ways and therefore they have encouraged Indian as well as foreign investors for establishment and span of their businesses. With the advent and span of East India Company of England in India, the British Raj was established. In the British era, the village panchayats were revived and again got full respect from the government. The British government has introduced the system of Zamindari for enhancing their access to each village and land revenue to be levied in terms of cash instead of food grains. The village assemblies or panchayats were become a media of levying taxes. It is evident from mentioned study that, concept of local government is not novel and it has its origin in ancient India [6].

In year 1687 the first Municipal Corporation in Madras was established without any legislative formulation with minimal powers of levying and collecting taxes on buildings [7]. The Charter Act of 1793 was the first legislative enactment for regulating municipal administration in presidency towns. The Bengal Act of 1842 was the first attempt to formulate municipal government outside the presidency town. In 1850 a new Act was enacted and according to that governments were allowed to create municipalities in their towns if required by its inhabitants. As the time passed the government felt that the involvement of local people is necessary to optimally utilize the public resources and finances. In this context the policy of financial decentralization was given by Lord Mayo in 1870 with the prime objective of involving Indians in administration as well as strengthening municipalities. The viceroy of India, Lord Ripon has avowed a momentous resolution on 18th May 1882 for reforming rural local self-government by augmenting local autonomy to municipalities. This resolution was also having intention to enhance the majority of elected representatives or non-officials in local bodies. Although, the resolution was having good organization and control; but despite of it was not prolific because the essence of democracy was lacking. As a result panchayats were dissolved and suppressed drastically and there was no institution through which a link between government and villages can be established. To overcome this problem, government appointed Royal Commission of decentralization in 1907. In 1909 the commission has given its report and a new era of local self-government was begun. The commission was also envisioned that local self-government should start from village level instead of district level. With a span of time, the government has analyzed that due to deceiving character and inadequate powers of these local bodies; they have not attained the intended success. In direction to remove these anomalies the commission emphasized on formulation of a genuine electorate comprises of members of village panchayat and representation of minorities through nominations. It has been recommended by the report of Indian Constitutional Reforms in 1918 that, the government should give direct attention towards development of panchayat system in villages. Finally, India got freedom on 15th August 1947 and at time India was having only three municipal corporations at Madras, Calcutta and Bombay with plenty of municipalities, town area committees, cantonment boards for cities and district boards for rural areas [8][9][10].

In 1947 Mahatma Gandhi has said that 'independence must begin at the bottom. Thus every village will be a republic or panchayat having full powers'. To ushering the aforesaid objectives, the Balwant Rai Mehta committee in 1957 recommended that Panchayati Raj Institutions (PRIs) should comprises of elected representatives and there should be sufficient autonomy, powers and freedom. But due to lack of management, insufficiency of funds and some political reasons, the growth of Panchayati Raj system was hindered. In 1977, the Ashok Mehta Committee was constituted for identifying key weaknesses of existing Panchayati Raj System. In August 1978, the committee has given their report with the recommendation of devising a system for supporting rural development and strengthening the planning process at grass root level. By the passage of time various other committees like Hanumantah Rao Working Group on Planning in year 1983, G.V.K. Rao Committee in year 1985 and Singhvi Committee in year 1986-1987 were constituted to triumph remedial solutions for strengthening Panchayati Raj system [11][12].

The current democratic architecture for local self-government is due to 73rd and 74th constitutional amendment Acts passed in 1992 for rural and urban areas respectively. According to 73rd constitutional amendment Act, organization of gram sabhas is mandatory and 3-tier architecture has to be established at zila, block and village level. All the posts at all the levels must be filled through elections (with two exceptions) and the minimum age for contesting will be 21 years. It also states that one third of the total seats must be reserved for women. Further, the tenure of a PRI will be of five years but if dissolves earlier then reconstruction of new PRI will also be through elections [13][14].

It is evident from above that the concept of local self-government is not new as it has its origin in ancient India. The Panchayati Raj Institutions were recommended as a tool for rural development therefore they need an apt management of their functioning and resources.

## 3. NEED OF ICT
India comprises of millions of people which are not able to fulfill even their basic needs. In such circumstances will it be rational to think about usage of ICT or adoption of electronic services with various objectives for such populace? However it seems peculiar but to think that ICT and e-governance can't be beneficial for such a nation where a huge population is not having accessibility to basic needs for their livelihood is a paradox. But it leads towards their economic and social development. ICT involves the representation of any information in digital form along with its electronic processing, storage, retrieval and dissemination. The information may comprise of news, circulars, reports, educational material, entertainment material and application forms etc. and can be accessed by many people in either horizontal or in sequential manner [15]. ICT has a capability to publicize any information to millions of people in minimal cost, time and efforts. The current era of globalization, marketization and increasing competitiveness requires that every citizen should be resourceful to run their livelihood enterprises. For seeking the solutions they have to be in contact with institutions irrespective of their location through electronic media. It can also expedite agricultural development and also beneficial in micro-finances administration. Internet facilitates people to interact with government, conduct businesses, communicate with peers, innovate, imbibe best practices into their lives and imitate their opinion [16]. Further, the access of ICT helps in creating sustainable economic relationships and efficient markets. Moreover, it is also helpful in eradication of poverty and improving health [17].

Globally there are various countries which are endeavoring towards rejuvenating their public administration by making it more proactive, accountable, service-oriented and transparent. This transformation requires intervention of technology in administration. And hence ICT can play significantly important role in advancement of public sector and its administration. Use of ICT in governance facilitates communications and enriches coordination of bureaucrats or elected representatives within different tiers of government [18].

In the current scenario only 40% of the rural populations have accessibility towards government services. Therefore, they need to be more empowered [19]. The current three-tier democratic architecture for rural areas intends towards participation of elected representatives in policy making. ICT enables the PRIs to proliferate their participation in governance and decision making processes by establishing communication between government and citizens. It also imparts transparency and accountability in the system. The spatial application like GIS (Geographical Information System) and application related to demographic data can be used for robust local planning [20][21].

## 4. NATIONAL POLICY ON INFORMATION TECHNOLOGY (NPIT)
This policy has been approved by Government of India on 14th September 2012 which visualizes India as a global IT hub by 2020. It also envisions the utilization of IT as an engine for rapid, inclusive and sustainable development of the Indian economy. It further envisages the use of ICT in all sectors and providing IT based solutions to citizen centric issues. There are two central objectives of the policy, firstly to bring full power of ICT and make it accessible to whole nation and secondly to harness human resources of whole India [22].

# 5. NATIONAL E-GOVERNANCE PLAN (NeGP)

E-governance is a new paradigm for governance by electronic means based upon the power of ICT. It imparts various types of e-services like G2C (Government to Citizen), G2B (Government to Business), G2G (Government to Government) and G2E (Government to Employee) [23]. The government of India had approved National e-Governance Plan on 18th May 2006 with a vision, "Make all Government services accessible to the common man in his locality, through common service delivery outlets, and ensure efficiency, transparency, and reliability of such services at affordable costs to realize the basic needs of the common man" [24]. It comprises of 31 Mission Mode Projects (MMPs) including central level MMPs, state level MMPs and local government level or integrated MMPs, where each MMP leads towards transforming a high priority citizen service from existing manual system to electronic system for delivering e-services. There are 11 central, 13 state and 7 integrated MMPs.

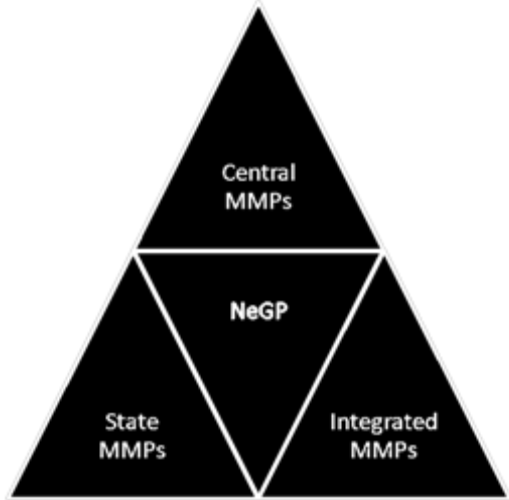

Fig 1 : National e-Governance Plan

**Table 1: List of MMPs under NeGP**

| Sr. | Central MMPs | State MMPs | Integrated MMps |
|---|---|---|---|
| 1. | Banking | PDS | CSC |
| 2. | Pensions | Education | e-Biz |
| 3. | Posts | Health | e-Courts |
| 4. | Passport | Employment Exchange | e-Procurement |
| 5. | UID | e-Panchayats | EDI |
| 6. | MCA21 | Agriculture | India Portal |
| 7. | IVFRT | CCTNS | NSDG |
| 8. | Insurance | Municipalities | |
| 9. | Income Tax | Treasuries | |
| 10. | e-Office | Commercial Taxes | |
| 11. | Central Excise | e-District | |
| 12. | | NRLMP | |
| 13. | | Road Transport | |

In this regard, the Government of India has also taken a loan of US $150 million from World Bank for improving the governance services pertaining to a common man [25]. Every MMP has its own objectives, scopes, implementation timeline and milestones along with measurable outcomes. The state governments can also define five MMPs according their own requirements.

## 5.1 Mechanism for service delivery

To impart variety of e-services, a robust service delivery mechanism has been devised comprises of various components mentioned as under:

### 5.1.1. State Wide Area Network (S.W.A.N.)

The objective behind SWAN is to connect all states/UT headquarters up to the Block level via District/ sub-Divisional Headquarters, in a vertical hierarchical structure with a minimum bandwidth capacity of 2 Mbps per link. Whereas the bandwidth can be enhanced up to 34 Mbps between state headquarter and district headquarter and 8 Mbps between district headquarter and block headquarter. To accomplish the objectives of SWAN, the government of India has sanctioned a total outlay of rupees 3,334 crore in year 2005 [26].

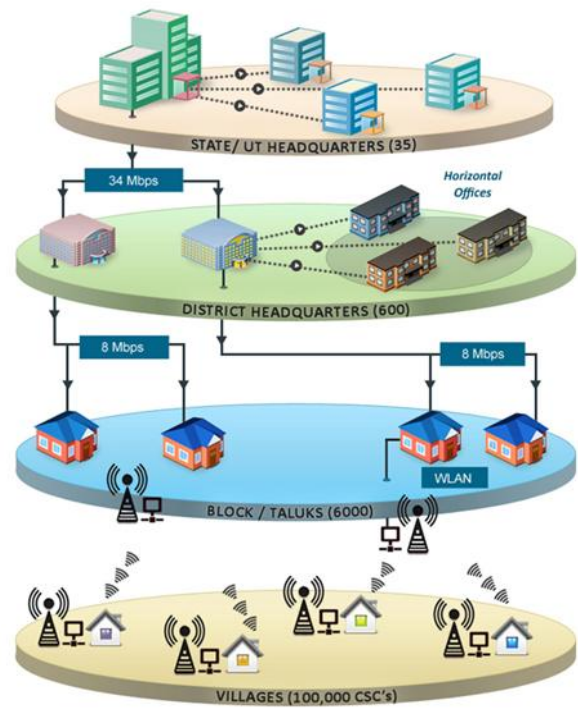

**Fig 2: State Wide Area Network**
**Source: www.negp.gov.in**

SWAN can be implemented by adopting any one of the two models i.e. NIC (National Informatics Center) model and PPP (Public Private Partnership) model. Currently, SWAN is functional in 28 states/ UT and in rest of the states it will be functional up to September, 2013 [27].

### 5.1.2. State Data Centre (SDC)

The government of India has approved SDC scheme with an outlay of rupees 1,623 crore and several associated objectives. According to this scheme each SDC will become central repository of state and provides secure storage of data, online delivery of services, citizen portal, state intranet portal, disaster recovery, service integration and remote management etc. Further, it also reduces the overall cost of data management and IT resources management [28]. Currently SDCs are operational in only 19 states/ UT and it is expected that they will be functional across the whole country by December, 2013 [29].

### 5.1.3. Common Service Centre (CSC)

The objective behind CSC scheme is to impart delivery of e-services at the grass root levels especially in rural areas. The services provided by CSCs are of high quality and cost effective in nature. They can provide information in the form of either audio, video or text in the areas of e-governance, education, health, telemedicine, entertainment as well as other private services [30]. The major services delivered by CSCs are e-district, land records, agricultural, domicile & character certificate, disbursal of social sector scheme benefits, birth & death certificates, property tax, electoral service, transport and grievances redressals etc. The implementation of the scheme is to be done by either NGOs (Non-Government Organizations) or PPPs. A CSC is having 3-tier architecture i.e. VLE (Village Level Operator), SCA (Service Center Agency) and SDA (State Designated Agency). Currently there are 96,411 rolled out CSCs across the whole country including all states and UTs whereas proposed number of CSCs are 1,26,071 [31].

### 5.1.4. National Service Delivery Gateway (NSDG)

It's a mission mode project with the objective of integrating information across all the departments in central, state and local government. Since government departments are having heterogeneous platforms and technologies therefore there integration is a mammoth task. It intends towards formulation of a middleware which can work as a core infrastructure for achieving standards based interoperability between various e-governance applications irrespective of their geographical locations and can reduce legacy investments in software & hardware [32].

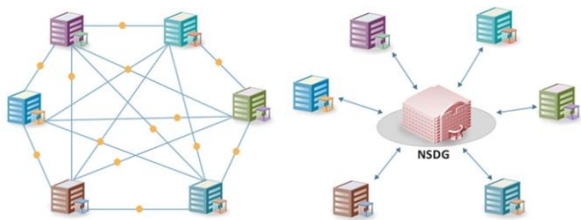

**Fig 3: National Service Delivery Gateway**
**Source: www.negp.gov.in**

## 6. LANDMARK PROJECTS FOCUSED TOWARDS RURAL REFORM

With the evolution of e-governance and NeGP, various e-governance applications were developed for automating innumerable segments of the administration and services across the whole country. This section discusses some key applications in the area of rural development.

## 6.1 e-Panchayat

This is a MMP which intends to improve quality of governance in PRIs which includes 0.235 million Gram Panchayats, 6094 Block Panchayats and 633 Zilla Panchayats. Further, it also enhances the coordination between Ministry of Panchayati Raj, Government of India and PRIs [33]. The central objective of this project is to ensure local area development and strengthen local self-governance by providing variety of services to its stakeholders. The stakeholders are elected representatives, Panchayat officials, citizens and other knowledge workers [34]. The Seventh Round Table Conference convened by Ministry of Panchayati Raj in 2004 had mentioned some major objectives in this regard that:

- It should be a decision support system (DSS) for the Panchayats
- A tool for improvement and internal management of Panchayats
- An efficient system for procurement
- A system for capacity building of elected representatives and bureaucrats
- A tool for enhancing transparency in the system

It has formed a base for the conceptualization e-Panchayats as a MMP with seven major intents mentioned in the figure 4.

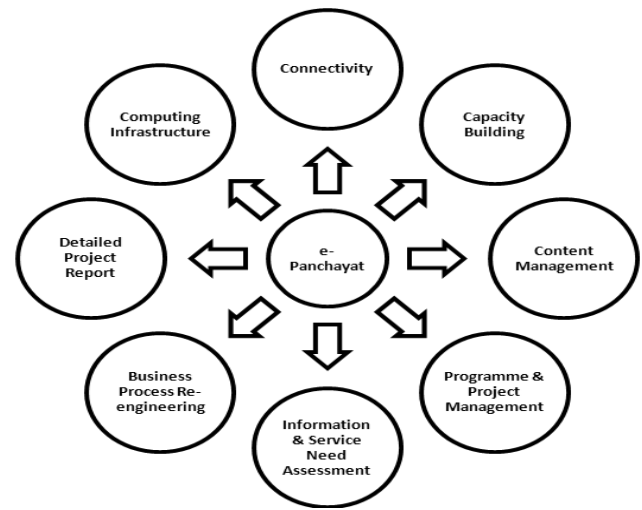

**Fig 4: Major objectives of e-Panchayat MMP**
**Source: www.epanchayat.gov.in**

After conceptualization of e-Panchayat; a sequence of study phases were started viz information & planning phase, information & service needs assessment phase, process re-engineering phase and detailed project report (DPR) preparation phase. Finally in 2009 e-Panchayat MMP was implemented by NIC with adequate capacity building mechanism and software operational manual.

### 6.1.1 Services Delivered by e-Panchayat

After finalization of project proposal and conceptualization of e-Panchayat, software development phase was initiated for the services intended to be delivered mentioned as under.

### 6.1.1.1 Local Government Directory (LGD)

The objective is to create and assign e-mail IDs to rural and urban local bodies along with management of their lists.

### 6.1.1.2 Area Profiler

This feature provides the socio-economic information, demographic details, details about public infrastructure and amenities etc.

### 6.1.1.3 Plan Plus

It facilitates PRIs in planning by assisting in data and process management associated with decentralized planning.

### 6.1.1.4 PRIAsoft

This application is focused towards management of financial accounts associated with various tiers of Panchayati Raj system.

### 6.1.1.5 ActionSoft

It facilitates PRIs in implementation and monitoring of various works or schemes.

### 6.1.1.6 National Asset Directory

It is an asset management systems which assist Panchayats in management of assets/ utilities created or owned by them.

### 6.1.1.7 Service Plus

The objective of this feature is to impart transparency in the system by managing grievances and their redressals.

### 6.1.1.8 Training/ Skill

This module manages the demands of training or skills among various levels of PRIs and also accomplishes their conduct.

### 6.1.1.9 Social Audit & Meeting Management

It manages the process of social audits conducted by Gram Sabha and their meetings.

### 6.1.1.10 National Panchayat Portal

It provides dynamic webpages to PRIs for content management and publishing.

### 6.1.1.11 GIS Layer

This module imparts spatial representation of some key reports like land records etc.

As it is a part of NeGP, therefore its regular monitoring will be conducted by NeGP apex committee which is headed by cabinet secretary, Government of India [35][36].

## 6.2 e-Mitra & e-Disha

In year 2002, the Government of Rajasthan initiated two G2C services viz LokMitra and JanMitra for urban and rural masses with the objective of delivering services related with various government departments to the citizens under a single umbrella. These two projects conquered the supposed objectives therefore, the Government of Rajasthan envisioned about expanding the scope of these projects. Hence e-Mitra project was initiated in 2004-2005 by integrating LokMitra and JanMitra together. Since it is implemented by adopting PPP approach therefore, besides delivery of numerous governmental services like bill/ tax collection, awareness generation and payment of service charges to various departments for generation of caste/ birth/ death certificates etc.; it also delivers various private sector services like insurance, loan, ATM (Automated Teller Machine), bill payments, credit cards, STD/ ISD, photocopy, fax, courier and internet café, hospitality etc. Figure 5 depicts the state's framework of e-Mitra [37].

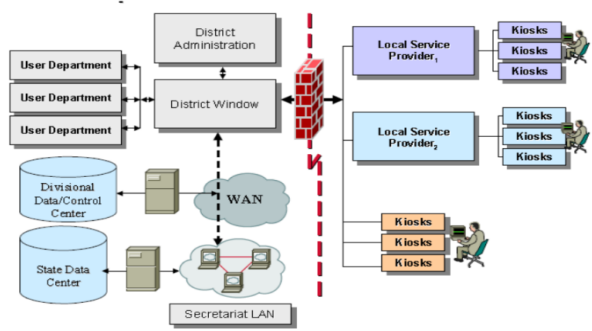

**Fig 5: State's Framework for e-Mitra**
**Source: http://www.cscmis.emitra.gov.in**

This project is functional in 33 districts of Rajasthan state and imparting services to the citizens at their doorsteps through CSCs or Kiosks on the basis of e-platform [38].

e-Disha is also a G2C application launched by Government of Haryana for ushering its citizens by imparting numerous services like issuance of various certificates, tax collection, water billing, social welfare schemes, arms license, learner & permanent driving license, vehicle registration, pension distribution, grievances redressals, and revenue records etc. The objective of project is to impart variety of services under a single roof with improved quality, enhanced transparency, real time availability, reduced costs, more citizen friendly environment and quick grievances redressals mechanism [39][40].

## 7. CONCLUSION

The concept of local self-governance is not novel but it has its existence with veracity even before Mauryan time. Although it was not having any legislative formulation despite of that it was functional with varying terminology. After independence with the recommendation of Balwant Rai Mahta committee; local self-governance especially in rural areas has attained constitutional recognition in the form of an Act. But in a country like India where 70% of the populace lives in rural areas with great diversification, the factors like rural connectivity, remoteness of geographical area etc. became major Impediments to accomplish desired objectives of decentralized governance. With the advent of ICT it has been taken a tool for dissemination of information. Finally in year 2006, the Government of India has formulated NeGP for automation of various mundane tasks and a significant attention will be given towards strengthening PRIs for improving local self-governance. In the current scenario various state governments have designed various applications for delivering services to citizens at their door step and ICT emerged as a tool for reinforcement of local self-governance.